\documentclass[11pt]{article}
\newcommand{\Comment}[1]{{}}
\usepackage{amsfonts,amsmath,epsfig,color,latexsym,graphics}
\usepackage{cite}
\Comment{\usepackage{color}
\definecolor{MyDarkBlue}{rgb}{0.15,0.15,0.45}
\usepackage[linktocpage=true]{hyperref}
\hypersetup{
colorlinks=true,
citecolor=MyDarkBlue,
linkcolor=MyDarkBlue,
urlcolor=MyDarkBlue,
pdfauthor={Stephen G. Naculich, Horatiu Nastase and Howard J. Schnitzer},
pdftitle={
},
pdfsubject={hep-th}
}

\usepackage[numbers,sort&compress]{natbib}
\usepackage{hypernat}}

\def\IZ{\relax\ifmmode\mathchoice
{\hbox{\cmss Z\kern-.4em Z}}{\hbox{\cmss Z\kern-.4em Z}}
{\lower.4pt\hbox{\cmsss Z\kern-.4em Z}}
{\lower1.2pt\hbox{\cmsss Z\kern-.4em Z}}\else{\cmss Z\kern-.4em Z}\fi}
\newcommand{\Z}{\mathsf{Z}\kern -5pt \mathsf{Z}}
\newcommand{\unit}{\mathsf{1}\kern -3pt \mathsf{l}}
\def\one{1\kern -3pt \mathrm{l}}

\def\ia{a}
\def\ib{b}
\def\ic{c}
\def\id{d}
\def\eps{\epsilon} 
\def\ep{\epsilon}
\def\bas{\bar{a}}
\def\as{a}
\def\mommu{ { s_{ij} \over \mu^2 } }
\def\mom{ { s_{ij} \over Q^2 } }
\def\Qmu{ {Q^2 \over \mu^2 } }
\def\bGam{\mathbf{\Gamma}}
\def\bS{{\bf S}}
\def\bT{{\bf T}}
\def\f{\tilde{f}}
\def\pel{{(\ell)}}
\def\Ell{{(L)}}

\def\EllL{{(L,L)}}
\def\Ellk{{(L,k)}}
\def\Ellodd{{(L,2m+1)}}
\def\Elleven{{(L,2m)}}
\def\Ellevenplustwo{{(L,2m+2)}}
\def\Ellzero{{(L,0)}}

\def\Elltwom{{(L,2m)}}
\def\Elltwomplus{{(L,2m+1)}}

\def\Zero{{(0)}}
\def\P{{(2P)}}
\def\NP{{(2NP)}}
\def\One{{(1)}}
\def\Onezero{{(1,0)}}
\def\Oneone{{(1,1)}}
\def\Two{{(2)}}
\def\Twozero{{(2,0)}}
\def\Twoone{{(2,1)}}
\def\Twotwo{{(2,2)}}

\def\ka {\kappa}
\def\grav{ {\Bigl( {\kappa \over 2} \Bigr)}}
\def \Tr {\mathop{\rm Tr}\nolimits}

\def\eqn#1{eq.~(\ref{#1})} 
\def\Eqn#1{Equation~(\ref{#1})}
\def\eqns#1#2{eqs.~(\ref{#1}) and~(\ref{#2})}

\def\theequation{\thesection.\arabic{equation}}

\newcommand\ignore[1]{}
\def\one{{\,\hbox{1\kern-.8mm l}}}

\def\ket#1{\left| #1\right\rangle}

\newcommand{\eg}{\emph{e.g.}\;}

\def\a{\alpha}\def\b{\beta}

\def\Z{\mathbb{Z}}

\newcommand{\Cset}{{\,\,{{{^{_{\pmb{\mid}}}}\kern-.45em{\mathrm C}}}}}

\newcommand{\cA}{\mathcal A}

\newcommand{\cI}{\mathcal I}

\newcommand{\cM}{\mathcal M}
\newcommand{\cN}{\mathcal N} \newcommand{\cO}{\mathcal O}

\newcommand{\cn}{{\cN}}

\newcommand{\be}{\begin{equation}}
\newcommand{\ba}{\begin{eqnarray}}
\newcommand{\bea}{\begin{eqnarray}}

\newcommand{\ee}{\end{equation}}
\newcommand{\eea}{\end{eqnarray}}
\newcommand{\ea}{\end{eqnarray}}

\newcommand{\nn}{\nonumber}

\def\eps{ \epsilon }

\begin{document}

\renewcommand{\thefootnote}{\fnsymbol{footnote}}

\makeatletter
\@addtoreset{equation}{section}
\makeatother
\renewcommand{\theequation}{\thesection.\arabic{equation}}

\rightline{}
\rightline{}
   \vspace{0.8truecm}

\begin{flushright}
BRX-TH-641\\ BOW-PH-153
\end{flushright}

\vspace{10pt}


\begin{center}
{\LARGE \bf{\sc 
Linear relations between $\cN \ge 4$ supergravity
\vspace{.1cm}

and subleading-color SYM amplitudes 
}}
\end{center} 
 \vspace{1truecm}
\thispagestyle{empty} \centerline{
{\large \bf {\sc Stephen G. Naculich${}^{a,}$}}\footnote{E-mail address: \Comment{\href{mailto:naculich@bowdoin.edu}}{\tt 
    naculich@bowdoin.edu}},
    {\large \bf {\sc Horatiu Nastase${}^{b,c,}$}}\footnote{E-mail address: \Comment{\href{mailto:nastase@ift.unesp.br}}{\tt 
    nastase@ift.unesp.br}}, {\bf{\sc and}}
    {\large \bf {\sc Howard J. Schnitzer${}^{c,}$}}\footnote{E-mail address:
                                \Comment{ \href{mailto:schnitzr@brandeis.edu}}{\tt schnitzr@brandeis.edu}}
                                                           }

\vspace{1cm}

\centerline{{\it$^{a}$ Department of Physics}} \centerline{{\it
Bowdoin College, Brunswick, ME 04011, USA}}

\vspace{.5cm}

\centerline{{\it ${}^b$ 
Instituto de F\'{i}sica Te\'{o}rica, UNESP-Universidade Estadual Paulista}} \centerline{{\it 
R. Dr. Bento T. Ferraz 271, Bl. II, Sao Paulo 01140-070, SP, Brazil}}

\vspace{.5cm}
\centerline{{\it ${}^c$ 
Theoretical Physics Group, Martin Fisher School of Physics}} \centerline{{\it Brandeis University, Waltham, 
MA 02454, USA}}

\vspace{1truecm}

\thispagestyle{empty}

\centerline{\sc Abstract}

\vspace{.1truecm}

\begin{center}
\begin{minipage}[c]{380pt}{\noindent 
The IR divergences of supergravity amplitudes 
are less severe than those of planar SYM amplitudes,
and are comparable to those subleading-color SYM amplitudes
that are most subleading in the $1/N$ expansion, 
namely $\cO(1/\eps^L)$ for $L$-loop amplitudes.
We derive linear relations between
one- and two-loop four-point amplitudes and one-loop
five-point amplitudes of $\cN =  4$, 5, and 6 supergravity 
and the most-subleading-color contributions of
the analogous amplitudes of $\cN= 0$, 1, and 2 SYM theory,
extending earlier results for $\cN=8$ supergravity amplitudes.
Our work relies on linear relations
between $\cN \ge 4$ supergravity 
and planar SYM amplitudes 
that were recently derived using the double-copy property of gravity,
and color-kinematic duality of gauge theories.
}
\end{minipage}
\end{center}

\vspace{.5cm}

\setcounter{page}{1}
\setcounter{tocdepth}{2}

\newpage

\renewcommand{\thefootnote}{\arabic{footnote}}
\setcounter{footnote}{0}

\linespread{1.1}
\parskip 4pt

\section{Introduction}

Recent progress in the understanding of
perturbative gauge theory and gravity amplitudes 
has been fueled by the discovery of a color-kinematic duality 
of gauge theory amplitudes by Bern, Carrasco, and 
Johansson \cite{Bern:2008qj,Bern:2010ue}.
Any $L$-loop $n$-point gauge theory scattering amplitude
where all particles are in the adjoint representation can be written as
\be
 \cA^\Ell_n = i^L \, g^{n - 2 + 2 L} \,
 \sum_{i}{\int \prod_{l=1}^L \frac{ d^{D} p_l}{(2 \pi)^{D}}
  \frac{1}{S_i}  \frac {n_i c_i}{\prod_{\alpha_i}{p^2_{\alpha_i}}}}\,.
\label{bcjloop}
\ee
The sum runs over a set of $L$-loop $n$-point diagrams
containing only cubic vertices,
the integral is over a set of $L$ loop momenta,
the product in the denominator contains all
the propagators of the diagram,  
the $n_i$ are kinematic numerators depending on momenta,
polarizations,  etc., and the $S_i$ are symmetry factors. 
The $c_i$ are color factors associated with the diagrams;
they are not all linearly independent, but satisfy various
(Jacobi) relations. 
Because of this, the representation (\ref{bcjloop}) 
of an amplitude is not unique;
different choices of kinematic numerators 
yield the same amplitude.
This freedom in the choice of $n_i$ is described as a generalized gauge 
transformation \cite{Bern:2010yg}.

A given representation (\ref{bcjloop}) is said
to satisfy color-kinematic (or BCJ) duality 
if the kinematic numerators $n_i$ obey precisely
the same set of algebraic relations observed by the color factors $c_i$.
It was conjectured that all gauge theories possess such a 
representation \cite{Bern:2008qj,Bern:2010ue},
and this was verified through three loops 
for the $\cN=4$ SYM four-point \cite{Bern:2010ue,Carrasco:2011hw}
and five-point \cite{Carrasco:2011mn} amplitudes.
It was further conjectured \cite{Bern:2010ue, Bern:2010yg}
that $L$-loop $n$-point gravity amplitudes 
can be constructed from a pair of gauge theory amplitudes as
\be
{\cM}_n^\Ell
\ =\ i^{L+1} \, \left( \frac{\ka}{2} \right)^{n - 2 + 2 L} \,
 \sum_i \int \prod_{l=1}^{L} \frac{d^Dp_l}{(2\pi)^D}
\frac{1}{S_i}\frac{n_i\tilde{n}_i}{\prod_{\alpha_i} p_{\alpha_i}^2} \,,
\label{doublecopy}
\ee
where $n_i$ and $\tilde{n}_i$  are the kinematic numerators
of the gauge theory representation (\ref{bcjloop}),
provided that at least one of the two 
representations satisfies color-kinematic duality.
The $\cN=8$ supergravity four-point \cite{Bern:2010ue,Carrasco:2011hw}
and five-point \cite{Carrasco:2011mn} amplitudes
were obtained as double copies (\ref{doublecopy})
of the corresponding $\cN=4$ SYM amplitudes.

The double-copy property of gravity theories was further used 
\cite{Bern:2011rj,BoucherVeronneau:2011qv} to obtain
new expressions for $\cN \ge 4$ supergravity amplitudes 
by combining loop amplitudes from an  
$\cN \le 4$ SYM theory  
(which need not satisfy BCJ duality) 
with the kinematic numerators from an $\cN=4$ SYM theory
(whose amplitudes do).
Specifically, it was shown in refs.~\cite{Bern:2011rj,BoucherVeronneau:2011qv} 
that the supergravity amplitudes on the left side of the following equation
\ba
\cN=8\; {\rm supergravity}&:&(\cN=4\; {\rm SYM})\times (\cN=4\;{\rm SYM})\cr
\cN=6\; {\rm supergravity}&:&(\cN=4\; {\rm SYM})\times (\cN=2\;{\rm SYM})\cr
\cN=5\; {\rm supergravity}&:&(\cN=4\; {\rm SYM})\times (\cN=1\;{\rm SYM})\cr
\cN=4\; {\rm supergravity}&:&(\cN=4\; {\rm SYM})\times (\cN=0\;{\rm SYM})
\label{mapping}
\ea
(where $\cN=0$ SYM denotes pure YM theory) 
can be represented as double copies of the gauge theories on the right
side, which have the same state multiplicities as 
the corresponding gravity theory.

Gravity amplitudes have IR divergences at loop level \cite{Weinberg:1965nx};
in dimensional regularization, 
the leading IR divergence of an $L$-loop amplitude goes as 
$\cO(1/\eps^L)$,   where $D=4- 2 \epsilon$ \cite{Dunbar:1995ed}.
Moreover, the IR-divergent terms of an $L$-loop gravity amplitude
are given by the exponentiation of the one-loop divergence
\cite{Weinberg:1965nx,Naculich:2011ry,White:2011yy,Akhoury:2011kq}. 
This behavior has been verified at two loops for $\cN=8$ supergravity 
\cite{Naculich:2008ew,Brandhuber:2008tf}
and also for $\cN \ge 4$ supergravity \cite{BoucherVeronneau:2011qv}.

The expressions  for various supergravity amplitudes obtained in 
refs.~\cite{Bern:1997nh,Carrasco:2011mn,Bern:2011rj,BoucherVeronneau:2011qv} 
involve combinations of planar SYM amplitudes, whose leading
IR divergence goes as  $\cO(1/\eps^{2L})$ at $L$ loops.
Therefore nontrivial cancellations among the SYM amplitudes are required to 
match the $\cO(1/\epsilon^L)$ leading IR behavior of the supergravity
amplitude.

IR divergences of gauge theory amplitudes that are subleading in the 
$1/N$ expansion are less severe than those of planar amplitudes.
In ref.~\cite{Naculich:2008ys,Naculich:2009cv} 
it was shown that the leading IR divergence of 
$A^\Ellk$, which denotes the subleading-color $L$-loop amplitude 
suppressed by $N^k$ relative to the planar amplitude,
is of $\cO(1/\eps^{2L-k})$ for $\cN=4$ SYM four-point amplitudes.
We will verify in sec.~\ref{sect-IR} that the same result
also holds for subleading-color $\cN<4$ SYM amplitudes.

The leading IR divergence of $A^\EllL$, the most-subleading-color $L$-loop
amplitude, therefore, is $\cO(1/\eps^L)$, precisely the same as the
leading divergence of the $L$-loop gravity amplitude.
One therefore speculates that such amplitudes could 
provide a basis for expanding gravity amplitudes. 
In previous work,  we showed that the one- and two-loop four-point 
functions \cite{Naculich:2008ys}
and the one-loop five-point function \cite{Naculich:2011fw} 
of $\cN=8$ supergravity are linearly related to the 
most-subleading-color amplitudes of $\cN=4$ SYM  theory.

In this paper, we show that analogous linear relations 
hold between certain amplitudes of $\cN + 4$ supergravity on the one hand, 
and the most-subleading-color amplitudes of $\cN$ SYM theory,
with $0 \le \cN < 4$, 
where the SYM theory is one of the two gauge theories 
in the double copy 
representation (\ref{mapping}).
We establish these relations by demonstrating their equivalence to the
expressions for $\cN \ge 4$ supergravity amplitudes
recently obtained in refs.~\cite{Bern:2011rj,BoucherVeronneau:2011qv}.

We point out that the linear supergravity/SYM relations 
between {\it integrated} amplitudes found in 
this paper and in refs.~\cite{Bern:1997nh,Naculich:2008ys,Carrasco:2011mn,Bern:2011rj,Naculich:2011fw,BoucherVeronneau:2011qv} 
involve amplitudes
in which the kinematic numerators are independent
of loop momenta, and so can be pulled outside the loop integrals.
It remains an open question whether similar relations 
between integrated SYM and supergravity amplitudes can
be obtained when the kinematic numerators depend on the loop momenta.

This paper is structured as follows. 
In section \ref{sect-leadingcolor}, we review the linear relations
between $\cN+4$ supergravity amplitudes and planar $\cN$ SYM amplitudes
obtained in refs.~\cite{Bern:2011rj,BoucherVeronneau:2011qv},
partly to establish notation.
In section \ref{sect-subleadingcolor},  we derive new linear relations
between $\cN+4$ supergravity amplitudes and subleading-color
$\cN$ SYM amplitudes, generalizing results previously obtained
in refs.~\cite{Naculich:2008ys,Naculich:2011fw}.
In section \ref{sect-IR},  the leading IR divergences of 
subleading-color amplitudes for generic SU($N$) gauge theories
are discussed,  and sec.~\ref{sect-concl} contains some concluding
remarks.

\section{Review of  relations between
supergravity and planar SYM amplitudes}
\label{sect-leadingcolor}

The color structure of a gauge theory amplitude may
be expressed by decomposing the amplitude 
in either a trace basis \cite{Bern:1990ux}
or a basis of color factors \cite{DelDuca:1999ha,DelDuca:1999rs}. 
The trace basis is more conducive to
exhibiting the $1/N$ expansion of the gauge theory,
while the color basis is more natural for 
exhibiting color-kinematic duality 
and for writing gravity amplitudes
as double copy of gauge theory amplitudes.
In this section, we will review the
trace basis for loop-level amplitudes, 
and then summarize the results 
derived in refs.~\cite{Bern:2011rj,BoucherVeronneau:2011qv}
for one- and two-loop four-point functions
and for one-loop five-point functions
of $\cN \ge 4$ supergravity.

\subsection{Trace basis}

One-loop $n$-point amplitudes of particles 
in the adjoint representation of an SU$(N)$ gauge theory 
may be decomposed into a basis of single and double traces \cite{Bern:1990ux}
\bea
{\cal A}^{(1)}(1,2,...,n)&=& g^n \sum_{j=1}^{\lfloor n/2 \rfloor+1}\sum_{\sigma\in S_n/S_{n;j}}
A_{n;j}(\sigma(1)...\sigma(n))
\, Gr_{n;j}(\sigma)
\nn \\
Gr_{n;1}(\one)&\equiv &N \, \Tr(T^{a_1}...T^{a_n})\nn \\ [1mm]
Gr_{n;j}(\one)&\equiv &\Tr(T^{a_1}...T^{a_{j-1}})\Tr(T^{a_j}...T^{a_n})
\label{nonplanar}
\eea
whose coefficients are referred to as color-ordered amplitudes.
Here $T^a$ are the generators in the defining representation of 
SU$(N)$, normalized according to $\Tr (T^a T^b) = \delta^{ab}$,
and $S_{n;j}$ is the subgroup of $S_n$ that leaves the 
trace structure $Gr_{n;j}$ invariant. 
At higher loops (for $n>5$),
the color decomposition will include 
triple and higher traces as well, 
and at $L$ loops, the amplitudes can contain powers of $N$ up to $N^L$, 
and so have an expansion in $1/N$.

Since in this paper
we focus primarily on 
four- and five-point functions in
supersymmetric Yang-Mills theories (with ${\cal N}$ supersymmetries),
we specialize the notation for these cases. 
For five-point functions, we have 
\ba
\cA^\One_\cn \, (1,2,3,4,5)
&=&
g^5 \bigg( \sum_{S_5/\Z_5 \times \Z_2} 
A^{\Onezero}_{12345, \, \cn} \, N \left[\Tr(12345) - \Tr(54321)\right]
\nn\\
&&
\hskip5mm
+ \sum_{S_5/\Z_2 \times S_3} 
A^{\Oneone}_{12;345, \, \cn} \, \Tr(12) \left[\Tr(345) - \Tr(543)\right]
\bigg)
\label{oneloopfivepointtrace}
\ea
where $A^\Onezero$ denotes the planar (leading-color) one-loop
amplitudes, $A^\Oneone$ the subleading-color one-loop amplitudes,
and $  \Tr(123\cdots) \equiv \Tr( T^{a_1} T^{a_2} T^{a_3} \cdots )  $.
Only single and double traces occur at any number of loops 
for four-point functions,  so we write
\ba
\cA^\Ell_\cn (1,2,3,4)
&=& g^{2 + 2L} 
\Big( 
A^{\Ell}_{1234, \,\cn}
\left[ \Tr(1234) + \Tr(1432)\right]
\nn\\
&&
\hskip0.9cm
+A^{\Ell}_{1342, \,\cn}
\left[ \Tr(1243) + \Tr(1342) \right]
\nn\\
&&
\hskip0.9cm
+A^{\Ell}_{1423, \,\cn}
\left[   \Tr(1423) + \Tr(1324) \right]
\nn\\
&&
\hskip0.9cm
+A^{\Ell}_{13;42,\,\cn}
\left[ \Tr(13) \Tr(24) \right]
\nn\\
&&
\hskip0.9cm
+A^{\Ell}_{14;23, \,\cn}
\left[\Tr(14) \Tr(23)   \right]
\nn\\
&&
\hskip0.9cm
+A^{\Ell}_{12;34, \,\cn}
\left[\Tr(12) \Tr(34)  \right]
\Big)
\label{4pttrace}
\ea
The color-ordered amplitudes $A^\Ell$ 
may be further decomposed in a $1/N$ expansion of amplitudes $A^\Ellk$, 
with $k=0, \cdots, L$.
Specifically, the amplitudes suppressed by even powers of $N$
relative to the leading-color (planar) amplitude
contribute to the single-trace amplitudes, and those
suppressed by odd powers of $N$
contribute to the double-trace amplitudes
\be
A^{\Ell}_{ijkl, \,\cn}
=
\sum_{m=0}^{\lfloor \frac{L}{2}  \rfloor} N^{L-2m} 
A^\Elltwom_{ijkl, \, \cn}, 
\qquad\qquad
A^{\Ell}_{ij;kl, \,\cn}
=
\sum_{m=0}^{\lfloor \frac{L-1}{2}  \rfloor}N^{L-2m-1} 
A^\Elltwomplus_{ij;kl, \, \cn}
\label{4pttracesub}
\ee

We will also find it useful to define ratios of loop-level to 
tree-level amplitudes for four-point functions as follows:
\be
M_{SYM,\;{\cN} }^\Elleven (s,t)
\equiv
\frac{  (g^2N)^L A_{1234,\cN}^\Elleven } 
       {A^\Zero_{1234}}, 
\qquad\qquad
M_{SYM,\;{\cN} }^\Ellodd (s,t)
\equiv
-\frac{(g^2N)^L  A_{13;42, \, \cN}^\Ellodd }{\sqrt{2}A^\Zero_{1234}}
\label{mratios}
\ee
where $s = (k_1 + k_2)^2$, $t=(k_1 + k_4)^2$ and $u=(k_1 + k_3)^2$
are the usual Mandelstam invariants,
with $k_i$ the momenta of the external particles,
and 
$A^\Zero_{1234}$ denotes the tree-level color-ordered amplitude,
which is independent of the number ${\cal N}$ of supersymmetries.

\subsection{Four-point amplitudes}

Alternatively, amplitudes may be decomposed in a basis 
of color factors \cite{DelDuca:1999ha,DelDuca:1999rs}. 
One-loop four-point amplitudes of particles in 
the adjoint representation of a gauge theory 
may be decomposed in a basis of color factors  of the 
box diagram  as
\be
\cA^{\One}_{\cn}\, (1,2,3,4) 
 =  g^4 \Big( c_{1234}\, A^{\Onezero}_{1234, \,\cn}
         + \, c_{1342}\, A^{\Onezero}_{1342, \,\cn}
          + \, c_{1423}\, A^{\Onezero}_{1423, \,\cn} \Big)
\label{4pt1loopampcolor}
\ee
where $c_{1234}$ is obtained by inserting a factor of 
the SU($N$) structure constants at each vertex of the box diagram
\be
c_{1234} = \f^{e a_1 b} \f^{b a_2 c} \f^{c a_3 d} \f^{d a_4 e}  \,.
\label{4pt1loopcolorfactor}
\ee
By using
\be
\f^{abc} =
i \sqrt2 f^{abc} = \Tr( [T^a, T^b] T^c )
\ee
to express $c_{1234}$ in terms of the trace basis (\ref{4pttrace}),
one ascertains \cite{DelDuca:1999rs}
that the coefficients in the one-loop 
color basis (\ref{4pt1loopampcolor}) are identical to
the planar one-loop color-ordered amplitudes 
$A^{\Onezero}_{ijkl, \,\cn}$
defined in \eqn{4pttracesub}.

Specializing to $\cN=4$ SYM, we write the coefficients in \eqn{4pt1loopampcolor}
in the form of \eqn{bcjloop}
\be
A^{\Onezero}_{1234, \,\cN=4} \, = \, i \, n_{1234} \, \cI^\One_{1234},
\qquad\quad
\cI^\One_{1234} = \int {d^D p \over (2\pi)^D}
\frac {1}{ p^2 \, (p - k_1)^2 \,(p - k_1 - k_2)^2 \, (p+k_4)^2}
\ee
where the kinematic numerators $n_{1234}$ (which in this case
are independent of the loop momentum and hence come outside the integral) 
satisfy
\be
n_{1234} = n_{1342} = n_{1423}  =  s t A^\Zero_{1234} \,.
\label{4pt1loopnum}
\ee
Because the numerator factors 
(\ref{4pt1loopnum}) for the $\cN=4$ SYM amplitude
satisfy color-kinematic duality,
they can be used in a double-copy 
representation (\ref{doublecopy}) of the one-loop four-point amplitude 
of $\cn+4$ supergravity   \cite{Bern:2011rj}
\ba
{\cM}^\One_{\cn+4} \, (1,2, 3,4)
 &=&
 i \grav^4 \Big( n_{1234}\, A^{\Onezero}_{1234, \,\cn}
         + \, n_{1342}\, A^{\Onezero}_{1342, \,\cn}
          + \,n_{1423}\, A^{\Onezero}_{1423, \,\cn} \Big)
\nn\\
&=&
\grav^4 
i s t A^\Zero_{1234}
\Big(A^{\Onezero}_{1234, \,\cn} + A^{\Onezero}_{1342, \,\cn}
          +A^{\Onezero}_{1423, \,\cn} \Big)
\label{4pt1loopampsg}
\ea
by replacing $g^4$ with $i (\kappa/2)^4$ 
and the color factors $c_{ijkl}$  with $n_{ijkl}$ 
in \eqn{4pt1loopampcolor}.
The $1/\eps^2$ poles of the planar amplitudes 
in \eqn{4pt1loopampsg}
cancel to give the expected $1/\eps$ leading IR divergence 
of one-loop supergravity amplitudes.

\begin{figure}[t]
\centerline{
{\epsfxsize4cm  \epsfbox{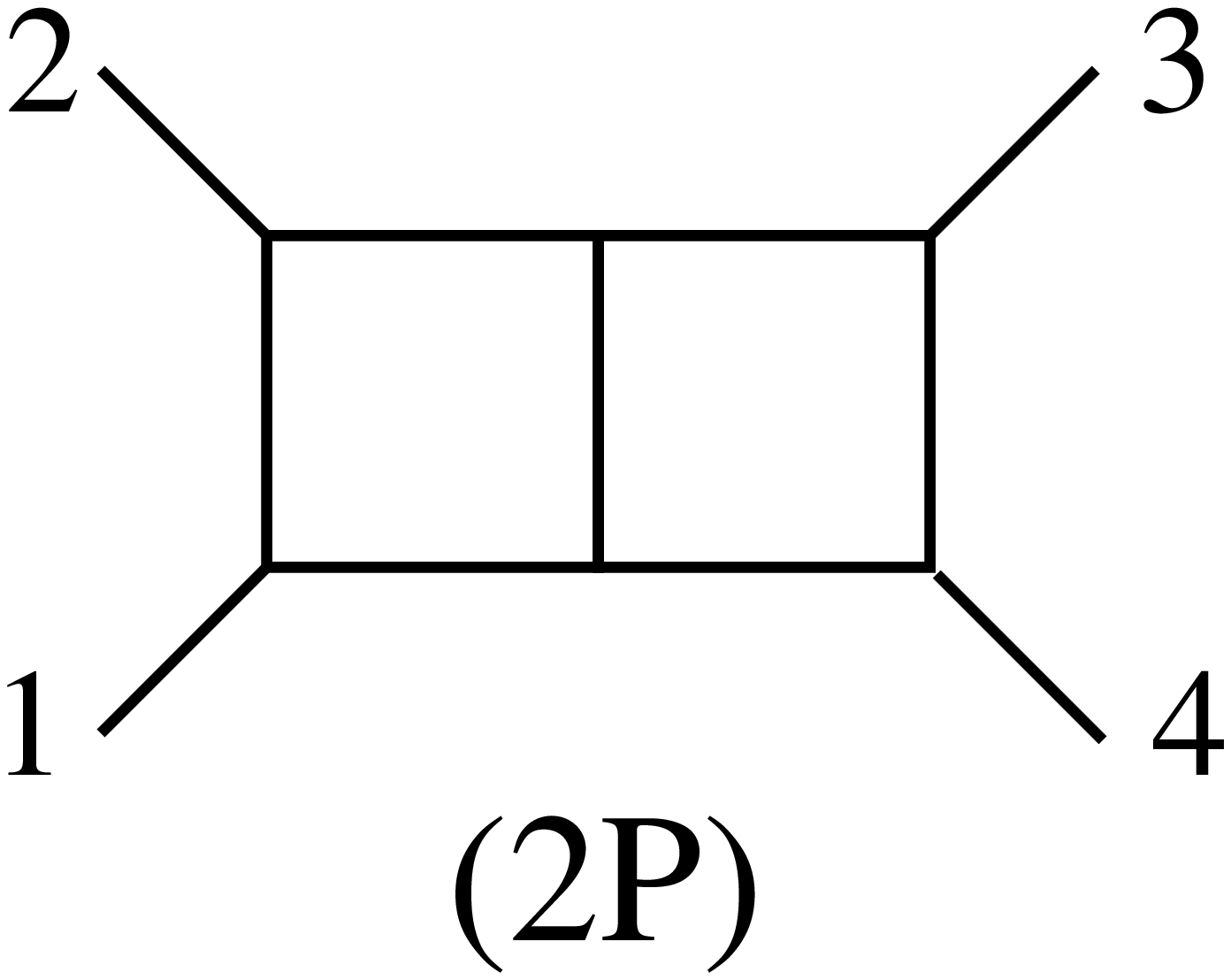}}
\hskip2cm
{\epsfxsize4cm  \epsfbox{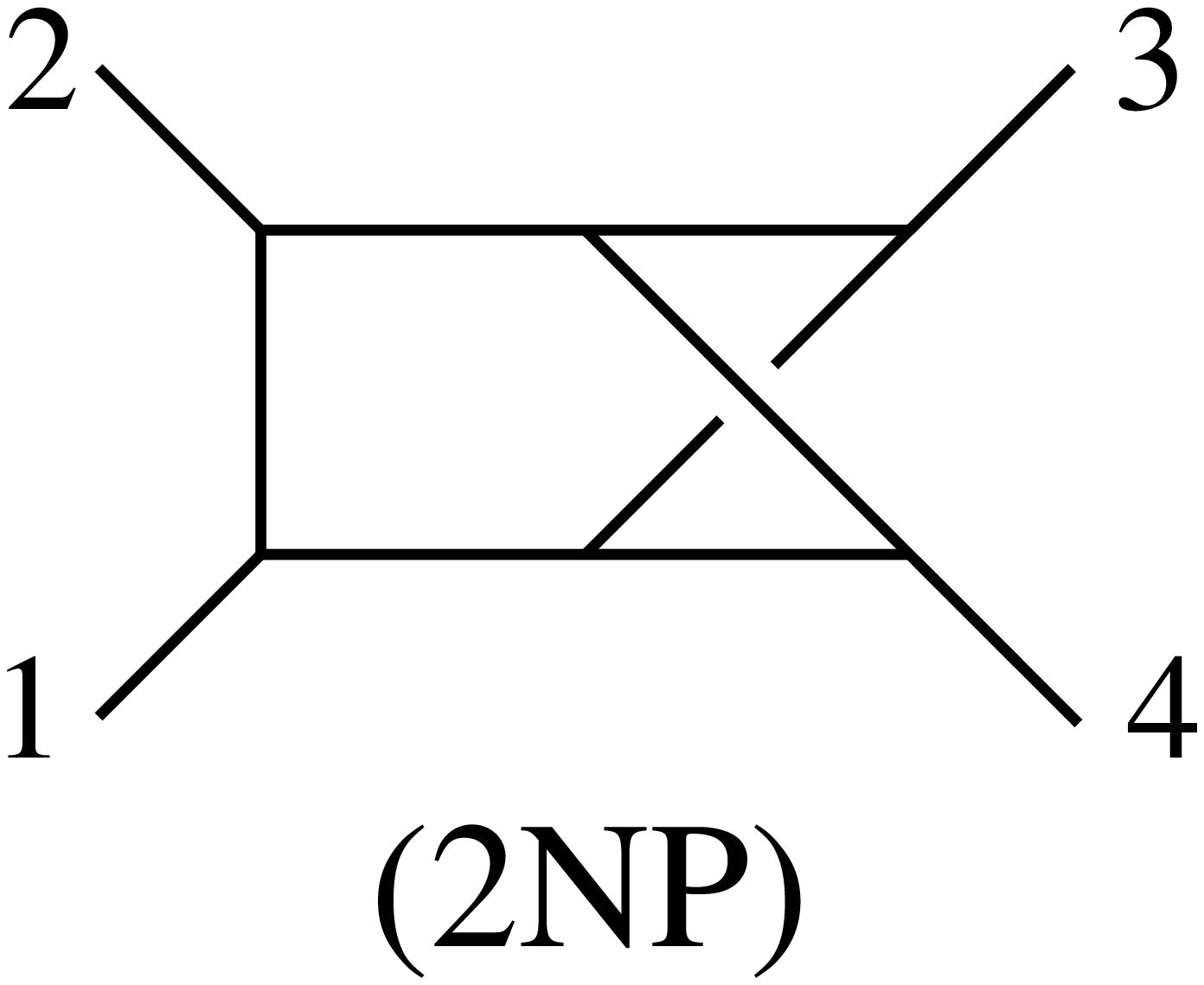}}
}
\caption{Two-loop planar and nonplanar diagrams} 
\label{fig-twoloop}
\end{figure}

Similarly, two-loop four-point amplitudes of particles
in the adjoint representation
may be expressed 
as\footnote{Our convention for $A^\P$ and $A^\NP$ differs 
from ref.~\cite{BoucherVeronneau:2011qv} by a sign.}
\ba
&& \hskip -0.9truecm 
\cA^{\Two}_{\cn}\, (1,2,3,4)
\nn \\
&& 
 = g^6 \Big( c^{\P}_{1234}\, A^{\P}_{1234, \,\cn}
            + c^{\P}_{3421}\, A^{\P}_{3421, \,\cn}
           + c^{\NP}_{1234}\, A^{\NP}_{1234, \,\cn}
           + c^{\NP}_{3421}\, A^{\NP}_{3421, \,\cn} \nn \\
&& \hskip .7cm
         + \, c^{\P}_{1342}\, A^{\P}_{1342, \,\cn}
            + c^{\P}_{4231}\, A^{\P}_{4231, \,\cn}
           + c^{\NP}_{1342}\, A^{\NP}_{1342, \,\cn}
           + c^{\NP}_{4231}\, A^{\NP}_{4231, \,\cn} 
\label{4pt2loopampcolor} \\
&& \hskip .7cm
          + \, c^{\P}_{1423}\, A^{\P}_{1423, \,\cn}
             + c^{\P}_{2341}\, A^{\P}_{2341, \,\cn}
            + c^{\NP}_{1423}\, A^{\NP}_{1423, \,\cn}
            + c^{\NP}_{2341}\, A^{\NP}_{2341, \,\cn} \Big) \,,
\nn
\ea
where the basis of color factors is given by the 
two-loop planar ($2P$) and nonplanar ($2NP$) diagrams 
shown in fig.~\ref{fig-twoloop}, namely,
\ba
c^{\P}_{1234}
&=& \f^{e a_1 b} \f^{b a_2 c} \f^{cgd} \f^{dfe} \f^{g a_3 h} \f^{h a_4 f} 
\label{4pt2loopplanar}
\\
c^{\NP}_{1234}
&=& \f^{e a_1 b} \f^{b a_2 c} \f^{cgd} \f^{hfe} \f^{g a_3 h} \f^{d a_4 f} 
\label{4pt2loopnonplanar}
\ea
The manifest symmetries
$c_{1234} = c_{4321} = c_{2143} = c_{3412}$
of both planar and nonplanar color factors have been used 
to reduce the number of terms in \eqn{4pt2loopampcolor}.
Even taking these symmetries into account, 
this color basis is still over-complete: 
the nonplanar color factors satisfy $c^\NP_{1234} = c^\NP_{1243}$,
and also can be rewritten in terms of planar ones
\be
3 c^\NP_{1234} =
c^\P_{1234} - c^\P_{2341} - c^\P_{1342} +  c^\P_{3421}  
\ee
and further there exists a linear relation among planar color factors
\be
0 = c^\P_{1234}- c^\P_{2341}  
  + c^\P_{1342}- c^\P_{3421}  
  + c^\P_{1423}- c^\P_{4231} 
\ee
so there are only five independent two-loop color factors. 
Representations of amplitudes that manifest BCJ duality, 
however, often require the use of an overcomplete color basis.

Specializing to $\cN=4$ SYM, \eqn{4pt2loopampcolor}
can be written in the form \eqn{bcjloop} with \cite{Bern:1997nh,Bern:1998ug}
\be
A^{\P}_{1234, \,\cN=4} \,=\, - \,   n^\P_{1234} \cI^\P_{1234},\qquad
A^{\NP}_{1234, \,\cN=4} \,=\, - \,   n^\NP_{1234} \cI^\NP_{1234},\qquad
\ee
where the two-loop scalar integrals associated with
the diagrams in fig.~\ref{fig-twoloop} are 
\ba
\cI^\P_{1234} &=& \int
 \frac{d^{D}p}{ (2\pi)^{D}} 
 \frac{d^{D}q}{ (2\pi)^{D}} 
\frac {1}{ p^2 \, (p - k_1)^2 \,(p - k_1 - k_2)^2 \,(p + q)^2 q^2 \,
        (q-k_4)^2 \, (q - k_3 - k_4)^2 } \,, \nn \\
\cI^\NP_{1234} &=& \int \frac{d^{D} p} { (2\pi)^{D}} \, 
           \frac {d^{D} q }{ (2\pi)^{D}} \
\frac{1}{ p^2\, (p-k_2)^2 \,(p+q)^2 \,(p+q+k_1)^2\,
  q^2 \, (q-k_3)^2 \, (q-k_3-k_4)^2}\, ,\nn
\ea
and the kinematic numerators (which are again independent of loop momenta)
are 
\ba
&&n^{\P}_{1234} = n^{\P}_{3421} = n^{\NP}_{1234} = n^{\NP}_{3421}=
 s ( s t A^\Zero_{1234})
\nn\\[1mm]
&&n^{\P}_{1342} = n^{\P}_{4231} = n^{\NP}_{1342} = n^{\NP}_{4231} =
 u ( s t A^\Zero_{1234})
\nn\\[1mm]
&&n^{\P}_{1423} = n^{\P}_{2341} = n^{\NP}_{1423} = n^{\NP}_{2341} =
 t ( s t A^\Zero_{1234})
\label{4pt2loopnum}
\ea
The numerators (\ref{4pt2loopnum}) satisfy color-kinematic duality
(see \eg the discussion in ref.~\cite{BoucherVeronneau:2011qv}),
and so they can be used in a double-copy representation of 
the two-loop four-point amplitude
of $\cn+4$ supergravity  \cite{BoucherVeronneau:2011qv}
\ba
&&
\cM^{\Two}_{\cn+4}\, (1,2,3,4)
 = i \grav^6  s t A^\Zero_{1234} 
\Bigg[ s \left( A^{\P}_{1234, \,\cn}
            + A^{\P}_{3421, \,\cn}
           + A^{\NP}_{1234, \,\cn}
           + A^{\NP}_{3421, \,\cn} \right)
\nn \\
&&
\hskip5.6cm
     + u \left( A^{\P}_{1342, \,\cn}
            + A^{\P}_{4231, \,\cn}
           + A^{\NP}_{1342, \,\cn}
           + A^{\NP}_{4231, \,\cn} 
\right)
\nn \\
&& 
\hskip5.6cm
     + t \left( A^{\P}_{1423, \,\cn}
             + A^{\P}_{2341, \,\cn}
            + A^{\NP}_{1423, \,\cn}
            + A^{\NP}_{2341, \,\cn} \right)
\Bigg]\hskip1cm 
\label{4pt2loopampsg}
\ea
obtained by replacing $g^6$ with $i (\kappa/2)^6$ 
and the color factors $c^{(2P,\,2NP)}_{ijkl}$  with $n^{(2P,\,2NP)}_{ijkl}$ 
in \eqn{4pt2loopampcolor}.
The $1/\eps^4$ and $1/\eps^3$ divergences
of the gauge theory amplitudes on the r.~h.~s.~cancel 
to give the expected $1/\epsilon^2$ leading IR divergence
for two-loop gravity amplitudes \cite{BoucherVeronneau:2011qv}.
Moreover, it was verified in ref.~\cite{BoucherVeronneau:2011qv}
that the IR divergent part of \eqn{4pt2loopampsg} 
is given by one-half of (the divergent part of)
the square of the one-loop supergravity amplitude \eqn{4pt1loopampsg}.

\subsection{Five-point amplitudes}

One-loop five-point amplitudes of particles in the adjoint
representation of a gauge theory 
may be decomposed in a basis of color factors of the 
pentagon diagram \cite{DelDuca:1999rs} 
as
\be
\cA^{\One}_{\cn}\, (1,2,3,4,5) 
=  g^5 \sum_{S_5/\Z_5 \times \Z_2}  
c^{(P)}_{12345}  A^\Onezero_{12345,\,\cn}
\label{5pt1loopampcolor} 
\ee
where
\be
c^{(P)}_{12345}=
\f^{g a_1 b} \f^{b a_2 c} \f^{c a_3 d} \f^{d a_4 e} \f^{e a_5 g} 
\label{5pt1loopcolorfactor}
\ee
is symmetric under cyclic permutations of $12345$ and 
antisymmetric under $12345 \to 54321$. 
By expressing $c^{(P)}_{12345}$ in terms of the trace 
basis (\ref{oneloopfivepointtrace}),
one ascertains \cite{DelDuca:1999rs,Naculich:2011fw}
that the coefficients in the decomposition (\ref{5pt1loopampcolor})
are identical to the planar one-loop color-ordered amplitudes.

Carrasco and Johansson \cite{Carrasco:2011mn} showed
that a representation of   the $\cN=4$ SYM five-point amplitude 
satisfying color-kinematic duality
requires the use of an overcomplete basis of color factors,
which includes the box-plus-line color diagram
\be
c^{(B)}_{12;345}
=
\f^{a_1 a_2 b} \f^{b  c g } \f^{c a_3 d} \f^{d a_4 e} \f^{e a_5 g} 
\label{boxplusline}
\ee
in addition to the pentagon color factors $c^{(P)}_{12345}$.
Then
\be
\cA^{\One}_{\cN=4}\, (1,2,3,4,5) 
=
  g^5 \sum_{S_5} \left( 
\frac{1}{10} i\,\beta_{12345} \, c^{(P)}_{12345} \,  \cI^{(P)}_{12345}
+
\frac{1}{4} i \, \gamma_{12;345}  \, c^{(B)}_{12;345}\,  \cI^{(B)}_{12;345}  \right)
\label{cjfirst}
\ee
where explicit expressions for the numerators $\beta_{12345}$ and $\gamma_{12;345}$ 
as well as the integrals $\cI^{(P)}_{12345}$ and $\cI^{(B)}_{12;345} $ 
are found in ref.~\cite{Carrasco:2011mn}.
For the purposes of sec.~\ref{sect-subleadingcolor}, 
we only need that 
$\beta_{12345}$ is symmetric under cyclic permutations of its indices
and antisymmetric under reversal of the indices,
and additionally satisfies the property  \cite{Carrasco:2011mn}
\be
\beta_{ijklm} - \beta_{jiklm} - \beta_{ijlkm} + \beta_{jilkm} =0 \,.
\label{betaproperty}
\ee
Because the kinematic numerators satisfy color-kinematic duality,
one may use them to obtain a double-copy representation 
of the two-loop five-point amplitude of $\cn+4$ 
supergravity  \cite{Bern:2011rj}
\be
\cM^{\One}_{\cn+4}\, (1,2,3,4,5) 
= \grav^5 
\sum_{S_5/(\Z_5 \times \Z_2)}  
i \beta _{12345}  A^\Onezero_{12345,\,\cn}
\label{5ptleadingrelation}
\ee
by replacing $g^5$ with $i (\kappa/2)^5$ 
and the color factors $c^{(P)}_{ijklm}$  with $\beta_{ijklm}$ 
in \eqn{5pt1loopampcolor}.

\section{Relations between supergravity and 
most-subleading-color SYM amplitudes}
\label{sect-subleadingcolor}

In the previous section, we reviewed various expressions
for $\cN \ge 4$ supergravity amplitudes in which
the expected leading $\cO(1/\epsilon^L)$ IR divergence 
emerges after cancellation of the higher-order IR divergences 
of planar gauge theory amplitudes.  
The IR divergences of gauge theory amplitudes that 
are subleading in the $1/N$ expansion
are less  severe than those of planar amplitudes
\cite{Naculich:2008ys,Naculich:2009cv,Nastase:2011mx},
as we will review in sec.~\ref{sect-IR},
with the most-subleading-color amplitude
having a leading divergence that is only $\cO(1/\eps^L)$.
This suggests the possibility of linear relations between
supergravity amplitudes and the most-subleading-color 
amplitudes of gauge theories used in the double copy 
representation.
In this section,
we will exhibit such relations for
one- and two-loop four-point functions
and the one-loop five-point function
of $\cN \ge 4$ supergravity.
In all these cases, the kinematic numerators of $\cN=4$ SYM theory 
are independent of the loop momenta.
Whether such relations exist more generally remains an open
question.

\subsection{Four-point amplitudes}

By expressing the four-point one-loop color factor
(\ref{4pt1loopcolorfactor}) in terms of the trace basis (\ref{4pttrace}),
one shows that the one-loop subleading-color amplitudes
are given by
\be
A^\Oneone_{13;42, \, \cn}  =
A^\Oneone_{14;23, \, \cn}  =
A^\Oneone_{12;34, \, \cn}  =
2 \left( A^\Onezero_{1234, \, \cn} +A^\Onezero_{1342, \, \cn} 
+A^\Onezero_{1423, \, \cn}  \right)
\label{4ptdecoupling}
\ee
which is just the one-loop U(1) decoupling 
relation \cite{Bern:1990ux}.
The relation (\ref{4ptdecoupling})
may be used to recast the one-loop $\cN+4$ supergravity amplitude
(\ref{4pt1loopampsg}) as
\be
{\cM}^\One_{\cn+4} \, (1,2, 3,4)
=
{1 \over 2} \grav^4 
i s t A^\Zero_{1234} A^{\Oneone}_{12;34, \,\cn} \label{onelooprela}
\ee
in terms of the one-loop subleading-color amplitude of $\cN$ SYM theory;
both sides have a leading $1/\eps$ IR divergence.
This generalizes the relation for the one-loop $\cN=8$ supergravity 
amplitude in ref.~\cite{Naculich:2008ys}.

Similarly, one may expand the 
two-loop color factors 
(\ref{4pt2loopplanar}) and (\ref{4pt2loopnonplanar})
in \eqn{4pt2loopampcolor} in terms of the trace basis
(\ref{4pttrace}), (\ref{4pttracesub}) 
to obtain \cite{Bern:1997nh}
\ba
A^\Twozero_{1234, \, \cn}  &=&
A^\P_{1234, \, \cn} +A^\P_{2341, \, \cn}
\\
A^\Twoone_{12;34, \, \cn}  &=&
2 \Big( 
3  A^{\P}_{1234, \,\cn}
+3 A^{\P}_{3421, \,\cn}
+2  A^{\NP}_{1234, \,\cn}
+2 A^{\NP}_{3421, \,\cn} \nn \\
&&- A^{\NP}_{1342, \,\cn}
- A^{\NP}_{4231, \,\cn} 
- A^{\NP}_{1423, \,\cn}
- A^{\NP}_{2341, \,\cn}\Big)
\\
A^\Twotwo_{1234, \, \cn}  &=&
2 \Big( 
 A^{\P}_{1234, \,\cn}
+ A^{\P}_{3421, \,\cn}
+ A^{\NP}_{1234, \,\cn}
+ A^{\NP}_{3421, \,\cn} \nn \\
&&
-2 A^{\P}_{1342, \,\cn}
-2 A^{\P}_{4231, \,\cn}
-2 A^{\NP}_{1342, \,\cn}
-2 A^{\NP}_{4231, \,\cn} \nn \\
&&
+ A^{\P}_{1423, \,\cn}
+ A^{\P}_{2341, \,\cn}
+ A^{\NP}_{1423, \,\cn}
+ A^{\NP}_{2341, \,\cn} \Big) \,.
\label{4pt2loopampsub}
\ea
As we show in sec.~\ref{sect-IR}, 
the leading IR pole of the amplitude $A^\Ellk$  goes as $\cO(1/\eps^{2L-k})$,
and so the leading IR divergence of $A^\Twotwo$ matches that of the two-loop
supergravity amplitude.   This is reflected in the following expression 
for the two-loop $\cn+4$ supergravity amplitude 
\be
\cM^{\Two}_{\cn+4}\, (1,2,3,4)
 =  {-\, {1 \over 6} } \grav^6   i s t A^\Zero_{1234} 
\left[ u A^\Twotwo_{1234, \, \cn} 
+ t A^\Twotwo_{1342, \, \cn} 
+s A^\Twotwo_{1423, \, \cn} 
\right]
\label{4pt2loopsgsub} 
\ee
which is easily verified by 
substituting \eqn{4pt2loopampsub} into \eqn{4pt2loopsgsub}
and comparing with \eqn{4pt2loopampsg}. 
\Eqn{4pt2loopsgsub}
generalizes the result in ref.~\cite{Naculich:2008ys}
expressing the two-loop $\cN=8$ supergravity amplitude
in terms of subleading-color $\cN=4$ SYM amplitudes.

Both one-loop (\ref{onelooprela})
and two-loop (\ref{4pt2loopsgsub}) relations 
may be written in a uniform way 
\be
\left(\sqrt{2}g^2N\right)^L M_{SG,\;\cN +4}^{(L)}(s,t)
= \frac{1}{3}
\left[ \left( \grav^2 u\right)^L  M_{SYM,\; \cN}^{(L,L)}(s,t) 
+ {\rm cyclic~perms~of~} s, t, u \right]
\label{uniform}
\ee
valid for $L=0,1,2$ and for ${\cal N}=0,1,2,4$, 
in terms of the ratios (\ref{mratios}) of $\cN$ SYM amplitudes and
the ratio of loop-level to tree-level $\cN +4$ supergravity amplitudes 
\be
M_{SG,\;{\cal N}+4}^{(L)}(s,t)
\equiv \frac{{\cal M}_{{\cal N}+4}^{(L)}(1,2,3,4)}{{\cal M}^\Zero(1,2,3,4)}
\ee
where the tree-level supergravity amplitude is given by the KLT relation 
\be 
{\cal M}^\Zero(1,2,3,4)=- \frac{ist}{u} [A_{1234}^\Zero]^2
\ee
independent of the number of supersymmetries.
This form
emphasizes that the most subleading SYM ratio $M_{SYM}^{(L,L)}$
is replaced by the supergravity ratio $M_{SG}^{(L)}$, 
and the dimensionless coupling of SYM 
$g^2N$ is replaced by the effective dimensionless supergravity coupling
$(\kappa/2)^2u$ (where the kinematic variable $u$ must 
appear for dimensional reasons).
Note that the simple form (\ref{uniform}) is not valid for $L > 2$ 
(though perhaps some modification thereof could be).

\subsection{Five-point amplitudes}

By evaluating the pentagon diagram color factor (\ref{5pt1loopcolorfactor})
in terms of the trace basis (\ref{oneloopfivepointtrace}),
one shows that the subleading-color amplitudes satisfy
\ba
 A^\Oneone_{12;345, \, \cn} 
&=&
A^\Onezero_{12345, \, \cn} 
+A^\Onezero_{23415, \, \cn} 
+A^\Onezero_{13425, \, \cn} 
+A^\Onezero_{34215, \, \cn} 
+A^\Onezero_{32415, \, \cn} 
+A^\Onezero_{13245, \, \cn} 
\nn\\
&&
\hskip-3mm
+A^\Onezero_{21345, \, \cn} 
+A^\Onezero_{23145, \, \cn} 
+A^\Onezero_{31425, \, \cn} 
+A^\Onezero_{34125, \, \cn} 
+A^\Onezero_{32145, \, \cn} 
+A^\Onezero_{31245, \, \cn}  \hskip1cm
\label{5ptdecoupling}
\ea
the relations written long ago in refs.~\cite{Bern:1990ux,Bern:1994zx}.
In sec.~\ref{sect-IR} we verify that the $1/\eps^2$ poles on the r.~h.~s.~of 
this equation cancel, leaving only an $\cO(1/\eps)$ pole for $A^\Oneone$.

Next, we will show that the following SYM-supergravity relation 
\be
\cM^{\One}_{\cn+4}\, (1,2,3,4,5) 
= 
 {1\over 20 } \grav^5
\sum_{S_5} i\beta_{12345}  A^\Oneone_{12;345, \, \cn}
\label{5ptsubleadingrelation}
\ee
which was proven in ref.~\cite{Naculich:2011fw} for $\cN=4$,
remains valid for $\cN \le 4$.
First, we use \eqn{5ptdecoupling} to rewrite \eqn{5ptsubleadingrelation} as
\be
\cM^{\One}_{\cn+4}\, (1,2,3,4,5) 
=\frac{1}{20}\left(\frac{\kappa}{2}\right)^5\sum_{S_5}
i\delta_{12345}
A^\Onezero_{12345, \, \cn}
\label{delta}
\ee
where 
\bea
\delta_{12345}&=&
\phantom{+}\b_{12345}+\b_{41235}+\b_{14235}+\b_{43125}+\b_{42135}+\b_{13245}\cr
&&+\b_{21345}+\b_{31245}+\b_{24135}+\b_{34125}+\b_{32145}+\b_{23145}
\eea
Here we have relabeled the sums over $S_5$ to give 
$A^\Onezero_{12345, \, \cn}$ as a common factor rather than  $\b_{12345}$.
We use the properties of $\beta_{12345}$
under cyclic permutation and reversal of indices to 
rewrite $\delta_{12345}$ as
\be
\delta_{12345}=\b_{12345}-\b_{21435}+\b_{14235}-\b_{41325}+\b_{13245}-\b_{31425}
\ee
Then we apply the property (\ref{betaproperty}) which holds 
for the kinematic numerator of the $\cN=4$  five-point amplitude
to obtain 
\be
\delta_{12345}=2\b_{12345}
\ee
Substituting this into \eqn{delta},
we see that \eqn{5ptsubleadingrelation} is equivalent 
to \eqn{5ptleadingrelation}.
Each of the terms on the r.h.s. of 
\eqn{5ptsubleadingrelation}
has a leading $\cO(1/\eps)$ divergence.

Therefore we again find a relation between 
supergravity and the most-subleading-color SYM amplitude. 
Note that its derivation depends not only on the group theory
relation (\ref{5ptdecoupling}), 
but also on properties (\ref{betaproperty})
that are valid because $\beta_{12345}$ 
are the kinematic numerators of the $\cN=4$ SYM theory,
which satisfy BCJ duality.

\section{IR divergences of subleading-color amplitudes}
\label{sect-IR}

In refs.~\cite{Naculich:2008ys,Naculich:2009cv,Nastase:2011mx} 
the structure of IR divergences of 
subleading-color amplitudes of $\cN=4$ theory was explored.
In this section, we will see that some of these properties 
carry over to $\cN < 4$ SYM theory.

\subsection{Four-point amplitudes}

We wish to show that the {\it leading} IR singularities of
the subleading-color amplitudes for 
four-point functions of particles in the adjoint representation
are the same in a generic SU($N$) gauge theory
as in $\cN=4$ SYM theory \cite{Naculich:2008ys,Naculich:2009cv}.
In particular, the color-ordered amplitude $A^\Ellk$ 
has leading pole of $\cO(1/\eps^{2L-k})$,
whose coefficient we will specify below.
Subleading IR poles, however, 
will differ for different gauge theories
due to the running of the coupling constant
and differing anomalous dimensions.

It is convenient \cite{Catani:1996jh,Catani:1998bh} 
to organize color-ordered amplitudes into a 
vector\footnote{We adopt here the basis 
used in ref.~\cite{Naculich:2011ep}, which differs slightly from 
than that used in refs.~\cite{Glover:2001af,Naculich:2009cv}.
Also, note that the overall factor of $g^2$ for a four-point function
$\cA (1,2,3,4)$
has been stripped off.}
\be
\ket{A}=
\left(
 A_{1234}, \, \, 
 A_{1342}, \, \, 
 A_{1423}, \, \, 
 A_{13;42}, \, \, 
 A_{14;23}, \, \, 
 A_{12;34}
\right) \,.
\label{vector}
\ee
We follow refs.~\cite{Sterman:2002qn,Aybat:2006mz}
by organizing the IR divergences as
\be
\label{factorize}
\left|   A \left(\mommu,  \as(\mu^2), \ep\right) \right> 
= 
 J \left(\Qmu, \as(\mu^2), \eps \right) \, 
{\bS} \left( \mom,\Qmu,  \as(\mu^2), \eps\right) 
\left | H \left( \mom,\Qmu,  \as(\mu^2), \eps \right) \right>
\ee
where the prefactors $J$ (``jet function") 
and ${\bS}$ (``soft function") characterize the 
long-distance IR-divergent behavior, 
while the short-distance behavior of the amplitude
is characterized by $\ket{H}$ (``hard function''),
and is finite as $\eps \to 0$.
(Quantities in boldface act as matrices on the color
space vectors.) 
In \eqn{factorize},
$s_{ij} = (k_i + k_j)^2$, 
$\mu$ is the  renormalization scale,
\be
a(\mu^2) = {g^2(\mu^2) N \over 8 \pi^2} (4 \pi e^{-\gamma} )^\eps 
\ee 
and $Q$ is an arbitrary factorization scale. 
The amplitude is independent of  the factorization scale,
but its split into $J$, $\bS$, and $H$ depends on $Q$. 

For $\cN=4$ SYM theory, which has $\beta_0= 0$, 
$\log J$ has only $1/\eps^2$ and $1/\eps$ poles, 
whose coefficients are given by anomalous dimensions.
For generic gauge theories, 
the coupling constant runs,
and so 
$\log J$ will also have higher-order poles,
starting at two-loop order in $\as$. 
These poles, whose coefficients depend on $\beta_0$, 
go up through $1/\eps^{L+1}$ at $L$-loop order 
\cite{Aybat:2006mz}. 
The one-loop $\cO(\as/\eps^2)$ contribution to $\log J$,
however, produces the leading $L$-loop IR divergence 
of $\cO(\as^L/\eps^{2L})$ in $J$, 
because the $\beta_0$-dependent contributions of $\cO(as^L/\eps^{L+1})$
are subleading in the $1/\eps$ expansion. 

The soft function 
\cite{Sterman:2002qn,Aybat:2006mz}
\be
{\bS} \left( \frac{s_{ij}}{Q^2}, \frac{Q^2}{\mu^2}, \as(\mu^2), \eps \right) 
\,=\,
{\rm P}~{\rm exp}\left[
\, -\; \frac{1}{2}\int_{0}^{Q^2} \frac{d\tilde{\mu}^2}{\tilde{\mu}^2}
\bGam \left( \frac{s_{ij}}{Q^2},
         \bas \left(\frac{\mu^2}{\tilde{\mu}^2}, \as(\mu^2), \eps  \right)
       \right) 
\right]\,
\label{soft}
\ee
depends on the anomalous dimension matrix 
\be
\bGam \left( \frac{s_{ij}}{Q^2}, \as(\mu^2) \right)
= \sum_{\ell=1}^\infty 
\as(\mu^2)^\ell \ \bGam^\pel
\left(\frac{s_{ij}}{Q^2}\right), 
\ee
where the leading form of the running coupling is given 
by \cite{Sterman:2002qn,Aybat:2006mz}
\be
{\bas}  \left(   \frac{\mu^2}{\tilde{\mu}^2}, \as(\mu^2), \eps \right) 
=  \as(\mu^2) \left( \frac{\mu^2}{\tilde{\mu}^2}   \right)^\eps  
\sum_{n=0}^\infty\left[\frac{\beta_0}{4\pi \epsilon}\left(
\left(     \frac{\mu^2}{\tilde{\mu}^2}   \right)^\eps-1\right)\as(\mu^2)\right]^n
\ee
If the matrices $\bGam^\pel$ all commute with one another
(which occurs if they are proportional to $\bGam^\One$, 
which is true through at least two loops \cite{Aybat:2006mz})
we can eliminate the path ordering of the exponential 
and integrate the terms to obtain 
\be
\bS \left( \frac{s_{ij}}{Q^2},\frac{Q^2}{\mu^2} , \as(\mu^2), \epsilon \right) 
 = \exp \left[ \frac{1}{2} \sum_{\ell=1}^\infty \as(\mu^2)^\ell
\left(\frac{\mu^2}{Q^2}\right)^{\ell\epsilon}
{ \bGam^{(l)}   \over  \ell \epsilon }
\left(1+ {\cO} \left(\frac{\as(\mu^2)}{\epsilon}\right)\right)
\right].
\ee
The omitted terms, which depend on $\beta_0$,
will be subleading
in the $1/\eps$ expansion relative to 
the $\bGam^\pel/\eps$ terms in $\log \bS$. 
Finally, when we exponentiate to obtain $\bS$,
we see that the contribution of $\bGam^\pel/\eps$ 
will be subleading in the $1/\eps$ expansion relative to the
one-loop contribution $\bGam^\One/\eps$,
as in the case of the jet function.

To summarize, we find,
as in the case of $\cN=4$ SYM theory \cite{Naculich:2009cv}, 
that the leading IR poles of the amplitude 
arise from the exponentiation of the 
one-loop  divergences\footnote{At this point, 
to simplify the formulas,
we set the factorization scale $Q$ equal to the renormalization scale $\mu$.}
\be
\ket{A} \Big|_{\rm most~divergent} = \exp \left(  - \frac{g^2 N}{4 \pi^2}
\left[\frac{1}{\ep^2} \one 
-\frac{1}{4\ep}\bGam^\One\right]\right)
\ket{A^\Zero}
\label{mostdiv}
\ee
We now decompose $\ket{A}$ in a loop and $1/N$ expansion
\be
\ket{A} = \sum_{L=0}^\infty \sum_{k=0}^L g^{2L} N^{L-k} \ket{A^\Ellk}
\ee
{}From \eqn{mostdiv},
the leading IR pole of the planar amplitude is simply
\be
\ket{A^\Ellzero } = 
\frac{1}{L!} \left(   -   \frac{1}{4 \pi^2}  \right)^L 
{1 \over\eps^{2L}}  
\ket{A^\Zero}
+ \cO\left( 1 \over \eps^{2L-1} \right) 
\ee
The one-loop anomalous dimension matrix is given by
\be
\bGam^\One = \frac{1}{N}  \sum_{i=1}^4 \sum_{j\neq i}^4 \bT_i \cdot \bT_j
 \log \left( {\mu^2 \over -s_{ij}  } \right) 
\ee 
where $\bT_i$ are the SU$(N)$ generators in 
the adjoint representation.
In the basis defined in \eqn{vector},
the anomalous dimension matrix takes the form 
\be
\bGam^\One 
=2 \left( 
\begin{array}{cc}
 \ia &     0 \\
    0  &  \id  
\end{array}
\right)
+ {2 \over N} \left( \begin{array}{cc} 0 & \ib \\
\ic & 0
\end{array}
\right)
\ee
with
\be
\ib = 
\left( 
\begin{array}{ccc}
0 & -Y & X \\ 
Z &  0 &- X \\ 
-Z&  Y & 0  
\end{array}
\right)
\qquad \qquad
\ic =
\left( 
\begin{array}{ccc}
0 & -2X & 2Y \\
2 X & 0 & - 2Z \\
-2Y & 2Z & 0
\end{array}
\right)
\ee
and
\be
X =\log \left(t \over u\right), \qquad
Y =\log \left(u \over s\right), \qquad
Z =\log \left(s \over t\right).
\ee
and the matrices $a$ and $d$ will not be needed.
{}From \eqn{mostdiv}, 
we can read off the coefficient of the leading IR pole 
for the subleading-color amplitudes
\be
\ket{A^\Ellk}=
\frac{1}{
k! (L-k)!  
(-4 \pi^2 )^L (-2)^k
\eps^{2L-k}
}
\left( 
\begin{array}{cc}
  0    & \ib   \\
 \ic   &  0
\end{array}
\right)^k \ket{A^\Zero}
+ \cO\left( 1 \over \eps^{2L-k-1} \right).
\ee
We can write this explicitly as 
\ba
\left.
\begin{array}{c}
A^\Ellodd_{13;42} \\
A^\Ellodd_{14;23} \\
A^\Ellodd_{12;34}
\end{array}
\right\}
&=&
\left[ 
(-1)^{L-1} 
\left( X^2 + Y^2 + Z^2 \right)^m 
\left(Y A^\Zero_{1423} - X A^\Zero_{1342} \right)
\over
(2m+1)! (L-2m-1)!  (4\pi^2)^L 2^m
\right]
{1 \over \eps^{2L-2m-1}}
\nn\\
&&+ 
\cO\left( 1 \over \eps^{2L-2m-2} \right)
\label{ellodd}
\ea
and
\ba
\left.
\begin{array}{c}
A^\Ellevenplustwo_{1234} \\
A^\Ellevenplustwo_{1342} \\
A^\Ellevenplustwo_{1423} 
\end{array}
\right\}
&=&
{
(-1)^{L} 
\left( X^2 + Y^2 + Z^2 \right)^m 
\left(Y A^\Zero_{1423} - X A^\Zero_{1342} \right)
\over 
(2m+2)! (L-2m-2)!
(4 \pi^2)^L 2^{m+1}
}  
\left\{
\begin{array}{c}
X-Y\\
Z-X\\
Y-Z
\end{array}
\right\}
{1 \over \eps^{2L-2m-2} }
\nn\\
&&+
\cO\left( 1 \over \eps^{2L-2m-3} \right)
\label{elleven}
\ea
where we observe that
the term in the numerator is invariant under cyclic permutations of 
$s$, $t$, and $u$:
\be
Y A^\Zero_{1423} - X A^\Zero_{1342} 
=Z A^\Zero_{1342}  - Y A^\Zero_{1234}
=X A^\Zero_{1234} -Z A^\Zero_{1423} 
\ee
as a consequence of the tree-level BCJ relations \cite{Bern:2008qj}
\be
t A^\Zero_{1423} = s A^\Zero_{1342} \, ,\qquad
u A^\Zero_{1342} = t A^\Zero_{1234}\, , \qquad
s A^\Zero_{1234} = u A^\Zero_{1423}  \,.
\ee
As expected, \eqns{ellodd}{elleven} are consistent with the group theory constraints on color-ordered
four-point amplitudes derived in ref.~\cite{Naculich:2011ep}.

\subsection{Five-point amplitudes and generalization to all $A_{n;j}$}

In ref.~\cite{Nastase:2011mx}, it was shown that the 
leading IR divergence of the subleading-color 
one-loop $n$-point amplitude 
is of $\cO(1/\eps)$ for $\cN=4$ SYM theory.
Here we show that the same result applies to generic
SU($N$) gauge theories.

First consider five-point functions.
We recall from \eqn{5ptdecoupling} 
that the one-loop subleading-color five-point amplitude 
obeys the relation
\ba
 A^\Oneone_{12;345} 
&=&
A^\Onezero_{12345} 
+A^\Onezero_{23415} 
+A^\Onezero_{13425} 
+A^\Onezero_{34215} 
+A^\Onezero_{32415} 
+A^\Onezero_{13245} 
\nn\\
&&
\hskip-3mm
+A^\Onezero_{21345} 
+A^\Onezero_{23145} 
+A^\Onezero_{31425} 
+A^\Onezero_{34125} 
+A^\Onezero_{32145} 
+A^\Onezero_{31245}  \hskip1cm
\label{5pt1loopdecoupling}
\ea
purely as a result of group theory.
The planar one-loop five-point amplitude can be written 
\be
A^\Onezero_{12345}  = A^\Zero_{12345}  M_{12345}
\label{prop}
\ee
For $\cN=4$ SYM theory, $M_{12345}$ is helicity-independent.
For generic SU($N$) gauge theories, this is no longer the case, 
except for the IR-divergent contribution \cite{Bern:1993mq}. 
In particular 
\be
M_{12345} = - \left( \frac{5}{16 \pi^2}  \right) \frac{1}{\eps^2}
+ \cO\left( 1 \over \eps \right)
\ee
Furthermore there are six relations among the twelve 
tree-level five-point amplitudes $A^\Zero_{12345}$,
known as Kleiss-Kuijf relations \cite{Kleiss:1988ne},
here shown for $n$-point tree amplitudes
\be
A^\Zero(1,\{\a\},n,\{\b\})=(-1)^{n_\b}\sum_{\{\sigma\}_i\in OP(\{\a\},\{\b^T\})}A^\Zero(1,\{\sigma\}_i,n)\label{KK}
\ee
It was shown in ref.~\cite{Nastase:2011mx} that 
these can be used to rewrite \eqns{5pt1loopdecoupling}{prop} as 
\bea
 A^\Oneone_{45;123} 
&=&
   A^\Onezero_{12345}[(M_{12345}-M_{41235})+(M_{43125}-M_{31245})]\nn\\[1mm]
&+&A^\Onezero_{12435}[(M_{12435}-M_{31245})+(M_{34125}-M_{41235})]\nn\\[1mm]
&+&A^\Onezero_{14235}[(M_{14235}-M_{31425})+(M_{34125}-M_{41235})]\nn\\[1mm]
&+&A^\Onezero_{13245}[(M_{23145}-M_{42315})+(M_{43125}-M_{31245})]\nn\\[1mm]
&+&A^\Onezero_{13425}[(M_{23145}-M_{31425})+(M_{43125}-M_{24315})]\nn\\[1mm]
&+&A^\Onezero_{14325}[(M_{23145}-M_{31425})+(M_{34125}-M_{23415})]
\label{5pointKK}
\eea
Due to the alternating signs in \eqn{5pointKK}, the
leading $\cO(1/\eps^2)$ pole cancels out, and 
$A^\Oneone_{12;345}$ goes as $\cO(1/\eps)$.

The same conclusion can be drawn for
arbitrary $n$-point functions of particles in the adjoint representation. Moreover, in that case, as we saw in (\ref{nonplanar}) we have a sum over 
double trace structures, multiplying color-ordered amplitudes $A_{n;j}$, and the same conclusion applies for all the $A_{n;j}$.
Since the ingredients of the proof rely only on group theory,
together with the fact that the leading IR divergence 
of the planar one-loop amplitude is
proportional to $1/\epsilon^2$ times the tree-level amplitude,
the alternating signs \cite{Nastase:2011mx}
in the one-loop decoupling relation
guarantee that the leading $1/\epsilon^2$ pole cancels out
of all the $A_{n;j}$ amplitudes
of a generic SU($N$) gauge theory,
generalizing the result of ref.~\cite{Nastase:2011mx}.

\subsection{Consistency of IR divergences for the five-point amplitude}

In this section we study the consequences of the relations in
sec.~\ref{sect-subleadingcolor} for the IR divergences of ${\cal N}$
SYM, and the consistency conditions associated with them.

In previous work by two of the authors \cite{Nastase:2010xa},
it was shown that in order to have a linear relation between 
the subleading-color five-point amplitude of ${\cal N}=4$ SYM 
and the five-point amplitude of ${\cal N}=8$ supergravity
\be
{\cal M}^{\rm (1)}_{{\cal N}=8}(1,2,3,4,5)
=\sum_{(ij)}\b_{(ij)}A^{(1,1)}_{ij;fgh, {\cal N}}
\label{linear}
\ee
where $(ij)$ are pairs of (different) indices, and $(ijfgh)$ is a permutation of $(12345)$,
the IR divergences must be the same, which gives 
\be
\frac{1}{\epsilon^2}{\cal M}^{(0)}(1,2,3,4,5)\sum_{i<j}s_{ij}(-s_{ij})^{-\epsilon}=\frac{r_\Gamma}{\epsilon^2}
\sum_{(fg)\neq lmn}\beta_{(fg)}\sum_{i<j}(-s_{ij})^{-\epsilon}\sum_{abc\in S_3}\epsilon_{lmn}[A^{(0)}_{ijabc}],\label{irdiv}
\ee
where the sum is over pairs $(fg)$ such that $(fglmn)$ is a permutation of $(12345)$, 
and the tree supergravity and SYM amplitudes 
are the same for all ${\cal N}$, and $\epsilon_{lmn}[A^{(0)}_{ijabc}]$ means we multiply the amplitude by the sign of the permutation 
$l,m,n$ inside $i,j,a,b,c$. In turn, for this relation to be true, the coefficients $\b_{(ij)}$ must satisfy the relation 
\be
\sum_{(fg)}N_{(ij),(fg)}\b_{(fg)}={\cal M}^{(0)}(1,2,3,4,5) s_{ij}\label{mustsat}
\ee
where 
\be
N_{(ij),(fg)}=\sum_{abc\in S_3}\epsilon_{lmn}[A^{(0)}_{ijabc}]
\ee
is a matrix of rank at most 9, since $\sum_{(ij)}N_{(ij),(fg)}=0$. Since the notation is a bit dense, we write a few examples of components 
for clarity
\bea
N_{(12),(12)}
&=&\sum_{perm.of(345)}\epsilon_{345}[A^{(0)}_{12(345)}]
=A^{(0)}_{12345}-A^{(0)}_{12543}-A^{(0)}_{12435}+A^{(0)}_{12534}-A^{(0)}_{12354}+A^{(0)}_{12453}\cr
N_{(12),(13)}
&=&\sum_{perm.of(345)}\epsilon_{245}[A^{(0)}_{12(345)}]
=A^{(0)}_{12345}-A^{(0)}_{12543}+A^{(0)}_{12435}-A^{(0)}_{12534}-A^{(0)}_{12354}+A^{(0)}_{12453}\cr
N_{(13),(13)}
&=&\sum_{perm.of(245)}\epsilon_{245}[A^{(0)}_{13(245)}]
=A^{(0)}_{13245}-A^{(0)}_{13542}-A^{(0)}_{13425}+A^{(0)}_{13524}-A^{(0)}_{13254}+A^{(0)}_{13452}
\cr
&&
\eea

Since by symmetry we only need to prove (\ref{mustsat}) for a single 
set of $(ij)$, the matching of the IR divergences implies
\be
{\cal M}^{(0)}(1,2,3,4,5) s_{12}=\sum_{(fg)}N_{(12),(fg)}\b_{(fg)}=\sum_{(fg)}\sum_{abc\in S_3}
\epsilon_{lmn}[A^{(0)}_{12abc}]\b_{(fg)}\label{needed}
\ee
In ref.~\cite{Nastase:2010xa}, 
the explicit solution for $\b_{(fg)}$ was not known.

We now show that the coefficients $\b_{(fg)}$ arising from the relation of the previous section do satisfy this relation. 
As shown in \cite{Naculich:2011fw}, we can rewrite 
\eqn{5ptsubleadingrelation} as
\be
{\cal M}_{{\cal N}=8}^{(1)}(1,2,3,4,5)
=i \left(\frac{\kappa}{2}\right)^5\sum_{S_5/{\mathbf Z}_2\times S_3}\frac{\gamma_{34}+\gamma_{45}+\gamma_{53}}{10}A^{(1,1)}_{12;345,{\cal N}=4}
\ee
where
\be
\gamma_{ij}=\b_{ijklm}-\b_{jiklm}
\ee
are independent of the order of $k,l,m$, are antisymmetric, $\gamma_{ij}=-\gamma_{ji}$ and satisfy
\be
\sum_{i=1}^5\gamma_{ij}=0
\ee
which means we have a basis of only 6 independent $\gamma_{ij}$'s, which we can take to be $\gamma_{12},\gamma_{13},\gamma_{14},\gamma_{23},\gamma_{24},
\gamma_{34}$.
We then obtain, by identifying with (\ref{linear}) that the $\beta_{(ij)}$ coefficients are given by
\be
\b_{(12)}=\frac{i}{10}(\gamma_{34}+\gamma_{45}+\gamma_{53})\label{beta}
\ee
and permutations. In the permutations, we have to be careful of the order in the sum, since $\gamma_{ij}=-\gamma_{ji}$. We take it to be the 
cyclic order of the remainder of the indices in the set $1,2,3,4,5$, since this is also the convention taken for the matrix $N_{(ij),(fg)}$.

Expanding the right hand side of (\ref{needed}) and substituting the coefficients from (\ref{beta}), and writing only in terms of the 
6 independent $\gamma_{ij}$ basis members, we obtain after some algebra
\bea
&\frac{i}{10}\Big\{&5A^{(0)}_{12345}[\gamma_{12}+\gamma_{13}+\gamma_{14}+\gamma_{23}+\gamma_{24}+\gamma_{34}]\cr
&+&5A^{(0)}_{12435}[\gamma_{12}+\gamma_{13}+\gamma_{14}+\gamma_{23}+\gamma_{24}-\gamma_{34}]\cr
&+&5A^{(0)}_{12453}[\gamma_{12}-\gamma_{13}+\gamma_{14}-\gamma_{23}+\gamma_{24}+\gamma_{34}]\cr
&+&5A^{(0)}_{12543}[\gamma_{12}-\gamma_{13}-\gamma_{14}-\gamma_{23}-\gamma_{24}-\gamma_{34}]\cr
&+&5A^{(0)}_{12534}[\gamma_{12}-\gamma_{13}-\gamma_{14}-\gamma_{23}-\gamma_{24}+\gamma_{34}]\cr
&+&5A^{(0)}_{12354}[\gamma_{12}+\gamma_{13}-\gamma_{14}+\gamma_{23}-\gamma_{24}-\gamma_{34}]\Big\}
\eea
which can then be finally written as 
\bea
{\cal M}^{(0)}(1,2,3,4,5)s_{12}&=&i\Big[A^{(0)}_{12345}\b_{12345}+A^{(0)}_{12435}\b_{12435}+A^{(0)}_{12543}\b_{12543}\cr
&&+A^{(0)}_{12534}\b_{12534}+A^{(0)}_{12453}\b_{12453}+A^{(0)}_{12354}\b_{12354}\Big]\cr
&&\label{tree}
\eea
or more generally
\be
{\cal M}^{(0)}(i,j,a,b,c)s_{ij}=\sum_{\sigma\in S_3}
i\,\b_{ij\sigma(a)\sigma(b)\sigma(c)}
\,A^{(0)}_{ij\sigma(a)\sigma(b)\sigma(c)}
\ee

This is a kind of tree level version of the BCJ relation 
(color-kinematic duality). 
However, note that the numerators are for 
{\em one-loop} amplitudes, yet they are used to multiply {\em tree amplitudes}.
This relation can be proven as follows. 

We take a specific helicity configuration, namely $(1^-2^-3^+4^+5^+)$. Then we have 
\bea
\b_{12345}&=&\langle 12\rangle^4 \frac{[12][23][34][45][51]}{\epsilon(1234)}\cr
A^{(0)}_{12345}&=&\frac{\langle 12\rangle^4}{\langle 12\rangle\langle 23\rangle\langle 34\rangle\langle 45\rangle\langle 51\rangle}\cr
{\cal M}^{(0)}(1,2,3,4,5)&=&i\frac{\langle 12\rangle^8\epsilon(1234)}{N(5)}\cr
N(5)&=&\prod_{i<j}\langle ij\rangle
\eea
and where the symbol $\epsilon(1234)$ satisfies
\be
\epsilon(1234)=4i\epsilon_{\mu\nu\rho\sigma}k_1^\mu k_2^\nu k_3^\rho k_4^\sigma =[12]\langle23\rangle[34]\langle41\rangle-\langle12\rangle[23]
\langle34\rangle[41],
\ee
is totally antisymmetric, and satisfies $\epsilon(1234)=-\epsilon(1235)$, etc., as can be easily checked.

We can then expand the right hand side of (\ref{tree}), and using the properties and form of $\epsilon(1234)$ we get after some algebra
\be
i\frac{\langle 12\rangle^8[12]}{N(5)}\Big[\langle14\rangle\langle24\rangle\langle35\rangle[34][45]+\langle13\rangle\langle
23\rangle\langle45\rangle[34][35]+\langle15\rangle\langle 25\rangle\langle 34\rangle[35][45]\Big]
\ee
Using the helicity spinor properties
\be
\langle ij\rangle\langle kl\rangle=\langle ik\rangle\langle jl\rangle +\langle il\rangle\langle kj\rangle, \qquad\qquad
\sum_{i\neq j,k}\langle ji\rangle[ik]=0
\ee
and $s_{12}=[12]\langle 21\rangle$, we finally find 
\be
i\frac{\langle 12\rangle^8s_{12}\epsilon(1234)}{N(5)}
\ee
which is the same as the left hand side of (\ref{tree}), 
therefore proving the relation.

We now review what we have done. We have explicitly proven equation (\ref{tree}), as a relation between ${\cal N}=4$ SYM and ${\cal N}=8$ 
supergravity, but since the objects involved are independent of ${\cal N}$, the relation is also true for SYM with ${\cal N}$ supersymmetries
and supergravity with ${\cal N}+4$ supersymmetries. 
We can trace that backwards however, to equation (\ref{needed}), since all the subsequent steps were independent of ${\cal N}$
(we only used group theory and the properties of the numerator coefficients $\b_{ijklm}$, which are still those of ${\cal N}=4$ SYM).
In turn, (\ref{needed}) came about from the consistency condition of matching the IR behaviors of (\ref{linear}), (\ref{irdiv}).
Since the left hand side of (\ref{linear}) is the known IR divergence of gravity amplitudes, still valid for ${\cal N}+4$ supergravity, we can 
consider it as a check that the IR divergence of $A^{(1,1)}_{fg;lmn,{\cal N}}$ is still given by 
\be
\frac{r_\Gamma}{\epsilon^2}\sum_{i<j}(-s_{ij})^{-\epsilon}\sum_{abc\neq i,j}\epsilon_{lmn}[A^{(0)}_{ijabc}]
\ee

\section{Conclusions}
\label{sect-concl}

In this paper, we have shown that various linear relations 
between amplitudes of ${\cal N}=8$ supergravity 
and the most-subleading-color amplitudes of 
${\cal N}=4$ SYM that were proven in 
refs.~\cite{Naculich:2008ys,Naculich:2011fw} 
remain valid for the analogous amplitudes of
${\cal N}+4$ supergravity and ${\cal N}$ SYM theory, 
for any ${\cal N}\leq 4$.
Specifically, the one- and two-loop four-point amplitudes 
of $\cN+4$ supergravity obey the relations
\ba
{\cM}^\One_{\cn+4} \, (1,2, 3,4)
&=&
{1 \over 2} \grav^4 
i s t A^\Zero_{1234} A^{\Oneone}_{12;34, \,\cn} 
\\
\cM^{\Two}_{\cn+4}\, (1,2,3,4)
 &=&  {-\, {1 \over 6} } \grav^6   i s t A^\Zero_{1234} 
\left[ u A^\Twotwo_{1234, \, \cn} 
+ t A^\Twotwo_{1342, \, \cn} 
+s A^\Twotwo_{1423, \, \cn} 
\right]
\ea
which together can be rewritten as
\be
\left(\sqrt{2}g^2N\right)^L M_{SG,\;\cN +4}^{(L)}(s,t)
= \frac{1}{3}
\left[ \left( \grav^2 u\right)^L  M_{SYM,\; \cN}^{(L,L)}(s,t) 
+ {\rm cyclic~perms~of~} s, t, u \right]
\ee
valid for $L=0,1,2$.
The one-loop five-point amplitudes satisfy
\be
\cM^{\One}_{\cn+4}\, (1,2,3,4,5) 
= 
 {1\over 20 } \grav^5
\sum_{S_5} i \beta_{12345}  A^\Oneone_{12;345, \, \cn}
\ee
We have also shown that the leading IR divergences 
of the subleading-color amplitudes 
$A^{(L,k)}$ for generic SU$(N)$ gauge theories 
are identical to those found for ${\cal N}=4 $ SYM 
theory \cite{Naculich:2008ys,Naculich:2009cv}.

We note that the results in this paper 
are valid for ${\cal N}=0$ YM, that is, pure glue theory, 
and therefore of possible interest for real world calculations. 
In particular, some of the results for IR divergences will 
generalize to QCD; this was already known to happen in some cases.

A central theme of this paper and our previous work on this subject is
the matching of IR divergences of supergravity to those of 
the most-subleading-color amplitudes of SYM theory. 
In detail, the BCJ color-kinematic duality has provided an
essential tool in proving some of these relations. 
It remains a challenge to exploit this point of view in cases 
when the BCJ numerators functions do not come outside of the integrals. 
Further generalizations to higher loops and/or $n$-points
are difficult, but we can hope they are still possible.

\section*{Acknowledgments}
\noindent 
The research of S.~Naculich is supported in part by the 
NSF under grant no.~PHY10-67961.
The research of H.~Nastase is supported in part by 
CNPQ grant 301219/2010-9 and by a grant from the SBF 
for travel to Brandeis University, 
where part of this project was carried out.
The research of H.~Schnitzer is supported in part by
the DOE under grant DE-FG02-92ER40706.

\vfil\break

\end{document}